\documentclass[12pt,english,aps, manuscript, a4paper, superscriptaddress]{revtex4}
\usepackage[T1]{fontenc}
\usepackage[latin9]{inputenc}
\usepackage{geometry}
\usepackage{float}
\usepackage{textcomp}
\usepackage{graphicx}
\usepackage{subscript}

\makeatletter
\@ifundefined{textcolor}{}
{%
 \definecolor{BLACK}{gray}{0}
 \definecolor{WHITE}{gray}{1}
 \definecolor{RED}{rgb}{1,0,0}
 \definecolor{GREEN}{rgb}{0,1,0}
 \definecolor{BLUE}{rgb}{0,0,1}
 \definecolor{CYAN}{cmyk}{1,0,0,0}
 \definecolor{MAGENTA}{cmyk}{0,1,0,0}
 \definecolor{YELLOW}{cmyk}{0,0,1,0}
}

\makeatother

\usepackage{babel}
\begin{document}
\title{Low Hole Effective Mass \emph{p}-type Transparent Conducting Oxides: Identification and Design Principles}
\author{Geoffroy Hautier}
\affiliation{Institut de la matière condensée et des nanosciences (IMCN), European Theoretical Spectroscopy Facility (ETSF), Université Catholique de Louvain, Chemin des étoiles 8, bte L7.03.01, 1348 Louvain-la-Neuve, Belgium}
\author{Anna Miglio}
\affiliation{Institut de la matière condensée et des nanosciences (IMCN), European Theoretical Spectroscopy Facility (ETSF), Université Catholique de Louvain, Chemin des étoiles 8, bte L7.03.01, 1348 Louvain-la-Neuve, Belgium}
\author{Gerbrand Ceder}
\affiliation{Department of Materials Science and Engineering, Massachusetts Institute of Technology, 77 Massachusetts Avenue, Cambridge, Massachusetts 02139, USA }
\author{Gian-Marco Rignanese}
\affiliation{Institut de la matière condensée et des nanosciences (IMCN), European Theoretical Spectroscopy Facility (ETSF), Université Catholique de Louvain, Chemin des étoiles 8, bte L7.03.01, 1348 Louvain-la-Neuve, Belgium}
\author{Xavier Gonze}
\affiliation{Institut de la matière condensée et des nanosciences (IMCN), European Theoretical Spectroscopy Facility (ETSF), Université Catholique de Louvain, Chemin des étoiles 8, bte L7.03.01, 1348 Louvain-la-Neuve, Belgium}

\begin{abstract}
The development of high performance transparent conducting oxides
(TCOs) is critical to many technologies from transparent electronics
to solar cells. While \textit{n}-type TCOs are present in many devices,
current \textit{p}-type TCOs are not largely commercialized as they
exhibit much lower carrier mobilities, due to the large hole effective
masses of most oxides. Here, we conduct a high-throughput computational
search on thousands of binary and ternary oxides and identify several
highly promising compounds displaying exceptionally low hole effective
masses (up to an order of magnitude lower than state of the art \textit{p}-type
TCOs) as well as wide band gaps. In addition to the discovery of specific
compounds, the chemical rationalization of our findings opens new
directions, beyond current Cu-based chemistries, for the design and
development of future \textit{p}-type TCOs.
\end{abstract}
\maketitle
Transparent conducting oxides (TCOs) are compounds exhibiting high
electrical conductivity and transparency to visible light. Those materials
are needed in many applications from solar cells, where a TCO thin
film provides electrical contact without impeding the flux of visible
light reaching the device, to transparent transistors that could,
for instance, be integrated in windows.\citep{MRS2000,Fortunato2012,Granqvist2007,Ellmer2012,Ginley2011}
The main strategy to achieve the two antagonistic properties of high
conductivity and transparency is to use wide band gap oxides (favoring
transparency) doped with a significant amount of mobile charge carriers,
either holes (\textit{p}-type) or electrons (\textit{n}-type).\citep{Edwards2004}\textit{
n}-TCOs (e.g., indium tin oxide, ITO) are already present in many
modern devices but \textit{p}-TCOs have not been largely commercialized
as their carrier mobilities stand an order of magnitude behind their
\textit{n}-counterparts. This situation impedes many critical technological
developments from more efficient organic and thin film solar cell
designs, benefiting from a better band matching by using \textit{p-}
instead of \textit{n}-TCOs,\citep{Liu2010,Beyer2007} to the entire
new field of transparent electronics which requires both \textit{p-}
and \textit{n}-type TCO materials.\citep{Kawazoe2000,Ohta2004,Wager2007,Sheng2006,Fortunato2012,Banerjee2005}

There is a fundamental reason to the difficulty of developing high
mobility \textit{p}-type TCOs: the localized oxygen \textit{p} nature
of the valence band in most oxides makes those bands very flat and
leads to large hole effective masses.\citep{Sheng2006,Banerjee2005}
The field of \textit{p}-type TCOs received most of its impulse a decade
ago when Kawazoe \textit{et al.} demonstrated that CuAlO\textsubscript{2}
delafossite could show encouraging \textit{p}-type conductivity and
optical transparency in the visible.\citep{Kawazoe1997} The unusual
hole mobility of CuAlO\textsubscript{2} was explained by a large
hybridization of the oxygen orbitals with 3\textit{d}\textsuperscript{10}
electrons in the Cu\textsuperscript{1+} closed shell, lowering the
oxygen character and leading to dispersive (low effective mass) valence
band. This finding led to the outline of a design rule for \textit{p}-type
TCOs requiring the presence of Cu\textsuperscript{1+} and motivated
the study of a very large range of Cu-based materials \citep{Banerjee2005,Sheng2006}
such as other delafossites (e.g., CuCrO\textsubscript{2},\citep{Nagarajan2001}),
SrCu\textsubscript{2}O\textsubscript{2}, \citep{Kudo1998,Ohta2002}
or Cu-based oxychalcogenides.\citep{Ueda2000} 

To this day, the question remains open wether alternative chemistries
and design rules could lead to materials with lower hole effective
mass. Answering this question is critical for the \textit{p}-type
TCO field as it would enable the identification of the high hole mobility
oxides that the TCO community has been looking for. Traditionally,
design principles are developped by the rationalization of experimentally
observed data. In this work, we take an alternative path using a database
of high-throughput \textit{ab initio} computed data containing electronic
structure for thousands of binary and ternary oxides.\citep{Service2012,Setyawan2010,Yang2012,Hautier2012}
By browsing this database, we identify the compounds and chemistries
leading to low hole effective masses. This enables us to uncover the
underlying chemical reasons for low hole effective masses and to propose
novel and unsuspected design rules for the development of \textit{p}-type
TCOs.

\section*{Results}

\subsection*{Hole vs electron effective mass distribution in oxides}

Our database contains density functional theory (DFT) band structures
for 3052 oxides. All the oxides that we have studied are \textit{existing}
minerals, or \textit{already synthetized} materials, whose experimentally
measured crystalline structure has been taken from the ICSD database.\citep{ICSD2006}
We have taken their first principles relaxed crystalline structure
as available in the Materials Project Database,\citep{MP,Jain} and
computed their electronic structure (band gaps and effective masses)
using state of the art methodologies, as described in the methods
section and in supplementary information. Figure \ref{fig:histogram-of-the}
shows the histogram of holes (in red) and electrons (in blue) effective
mass. The difference in distribution between hole and electron effective
mass compounds is striking, emphasizing that finding high mobility
\textit{p}-type oxides is indeed significantly more challenging than
for \textit{n}-types. The chemical reasons for such a difference comes
from the very different character of the valence and conduction bands
in oxides. The valence bands tend to be of oxygen \textit{p} localized
character (leading to large effective masses), while the conduction
bands are cationic and present more often dispersive bands (i.e.,
low effective mass).\citep{Minami1999a,Medvedeva2010}

\begin{figure}[H]
\begin{centering}
\includegraphics[width=15cm]{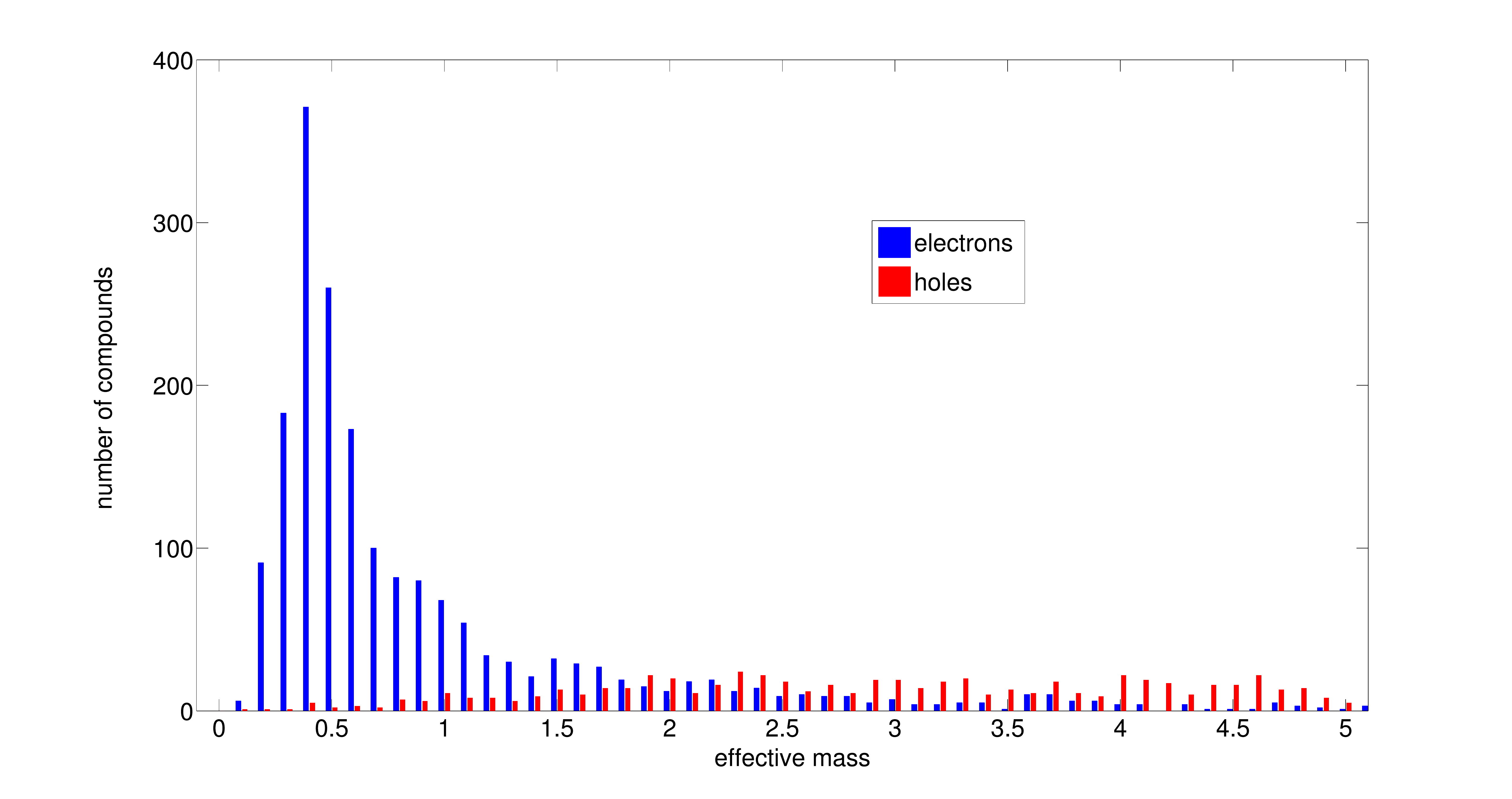}
\par\end{centering}

\caption{Histogram of the maximum line effective mass for holes (valence band)
in red and electrons (conduction band) in blue in our set of binary
and ternary oxides. The effective mass bin size is 0.2 and the figure
focuses on the region of low effective mass (lower than 5). \label{fig:histogram-of-the}}
\end{figure}

\subsection*{High-throughput identification of low hole effective mass, wide band
gap oxides}

\begin{figure}[H]
\begin{raggedright}
\includegraphics[width=18cm]{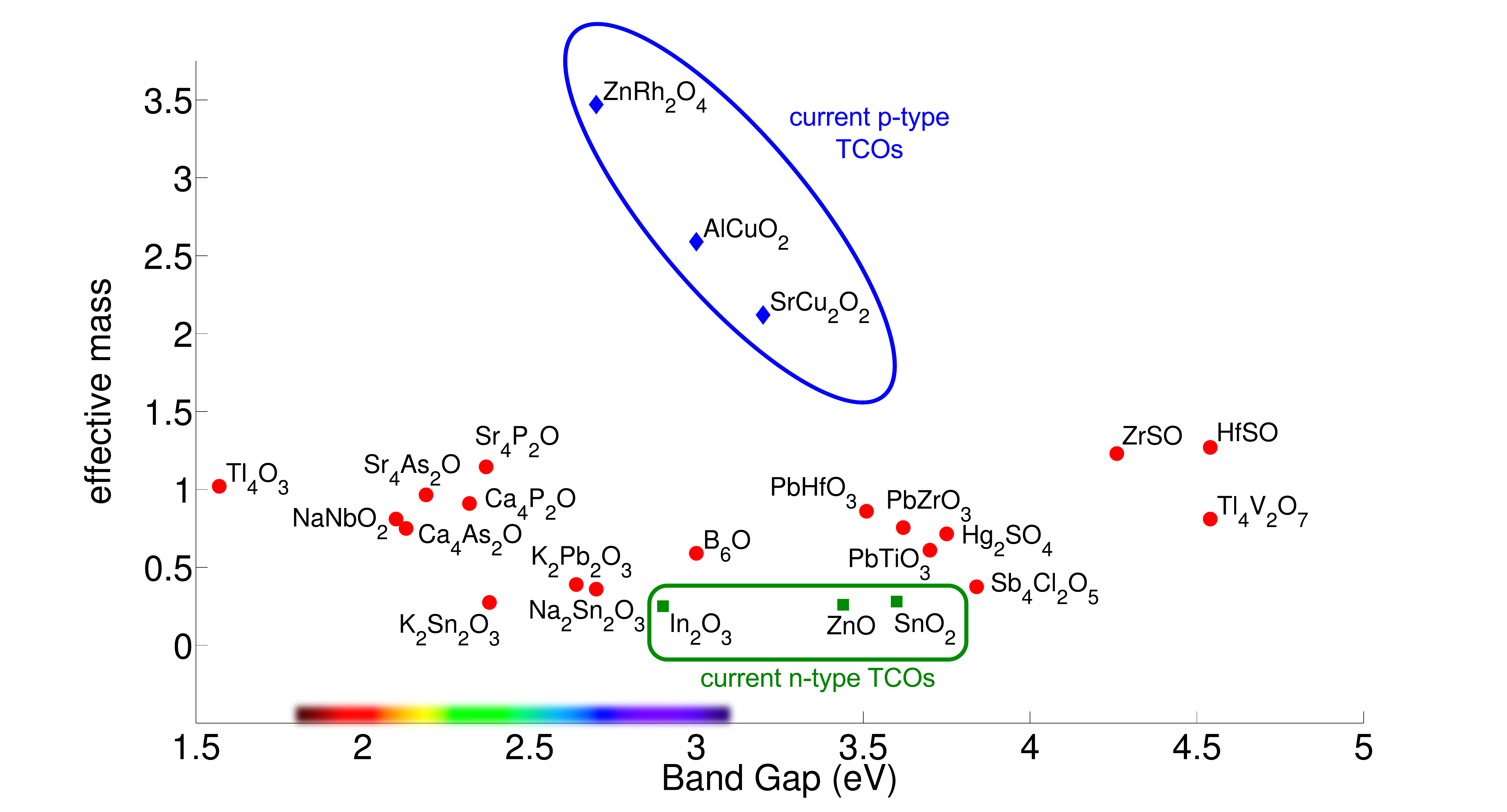}
\par\end{raggedright}

\caption{Effective mass vs band gap for the \textit{p}-type TCO candidates
(red dots). We superposed on the band gap axis a color spectrum corresponding
to the wavelength associated with a photon energy. A few known \textit{p}-type
(blue diamonds) and \textit{n}-type (green square) TCOs can be compared
to the new candidates. The best TCOs should lie in the lower right
corner. \label{fig:Effective-mass-vs}}
\end{figure}

While low hole effective mass oxides are rare, our large database
gives us the opportunity to identify the outliers. By filtering for
oxides with an effective mass lower than 1.5, we end up with only
20 candidates that have never been considered as TCOs. Details on
the screening procedure and results are available in methods and supplementary
information. To further assess the practical interest of those candidates,
we also evaluate their band gaps as this would influence their light
absorption. As DFT is known to not as accurately model band gaps than
band shape and width, we computed band gaps with a higher order method:
many body perturbation theory in the \textit{GW} approach.\citep{Hybertsen1986,Aulbur1999}
Figure \ref{fig:Effective-mass-vs} plots the effective mass vs the
band gap for the 20 identified compounds (red dots). For comparison,
we also plotted a few known \textit{p}-type TCOs (blue diamonds, Cu-based
TCOs and ZnRh\textsubscript{2}O\textsubscript{4}) and some current
\textit{n}-type materials (green squares, SnO\textsubscript{2}, In\textsubscript{2}O\textsubscript{3},
ZnO). The compounds we identified beat the state of the art \textit{p}-type
TCOs by up to an order of magnitude in effective mass. Some compounds
are even reaching hole effective mass values close to the best electron
effective mass in oxides exhibited by current \textit{n}-type materials.
This shows that the large difference in mobility between current \textit{n}-type
and \textit{p}-type materials is not inevitable and could be overcome
by the investigation of alternative chemistries. 

The best TCOs should lie in the lower right corner as they would show
small effective masses and large band gaps. However, the materials
containing toxic elements such as Pb, Tl and Hg are less likely to
be of interest technologically. The series of \textit{A}\textsubscript{4}\textit{B}\textsubscript{2}O
oxypicnitides compounds (\textit{A} = Ca, Sr and \textit{B} = P, As)
and NaNbO\textsubscript{2} are not of high priority as compounds
with both higher band gaps and lower effective masses are present
in our data set. K\textsubscript{2}Sn\textsubscript{2}O\textsubscript{3}
shows the lowest effective mass around 0.27-0.28 but a band gap on
the small side (2.4 eV). Interestingly, the band gap of K\textsubscript{2}Sn\textsubscript{2}O\textsubscript{3}
can be increased by substituting Na which would lead to lower absorption
in the visible (see supplementary information). On the other hand,
the ZrOS and HfOS compounds have larger effective masses but with
a significantly larger band gap and opportunities for full visible
range transmission. Finally, both Sb\textsubscript{4}Cl\textsubscript{2}O\textsubscript{5}
and B\textsubscript{6}O stand as the compounds of major interest
with both large band gaps (respectively 3.6 and 3 eV) and low effective
masses (respectively 0.37 and 0.59), while made of non-toxic, abundant
elements.

\subsection*{p-type dopability in the low hole effective mass, wide band gap candidates}

Effective masses and band gaps are necessary conditions for good p-type
TCOs but they are not sufficient. One key requirement for a p-type
TCO is to be able to be doped (intrinsically or extrinsically) with
acceptor defects that will generate holes in the valence band. It
is well reported that certain oxides are intrinsically difficult to
dope p. The easy formation of compensating intrinsic defects (hole-killers)
as the oxygen vacancy when lowering the Fermi energy towards the valence
has been a main issue for the p-doping of certain oxides. Doping studies
for other applications than TCOs have already been reported for several
of the chemistries we identified. B\textsubscript{6}O has been experimentally
measured to be p-type, PbTiO\textsubscript{3} has been identified
as p-type dopable by computations, in agreement with experimental
results on PbZr\textsubscript{0.5}Ti\textsubscript{0.5}O\textsubscript{3}
and recently computations showed that to oxygen vacancy is not a hole-killer
in ZrOS. For the remaining compounds of greatest interest, we performed
defect computations for all the vacancy intrinsic defects (see supplementary
information). From these results we can conclude that Sb\textsubscript{4}Cl\textsubscript{2}O\textsubscript{5}
is not likely to be p-type doped due to strong hole compensation by
oxygen and chlorine vacancies. On the other hand, K2Pb2O3 as well
as K2Sn2O3 (in both polymorphs) shows an oxygen vacancy very high
in energy and that will not compensate hole formation. Moreover, the
presence of a low energy potassium vacancy could lead to intrinsic
p-type behavior for those materials.

\subsection*{Discussion}

\begin{figure}[H]
\begin{centering}
\includegraphics[width=12cm]{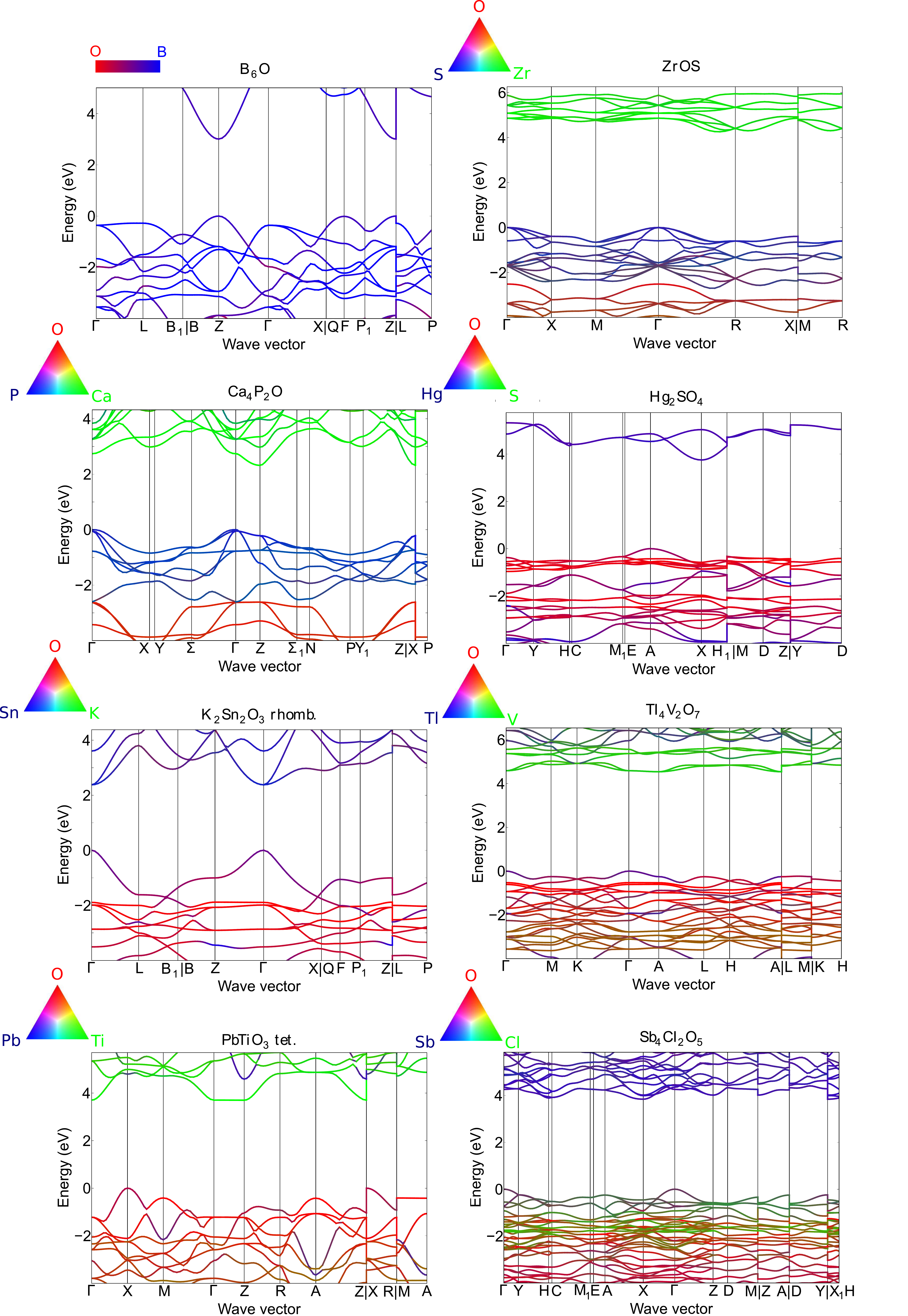}
\par\end{centering}

\caption{Band structures for a series of representative \textit{p}-type TCOs
identified in this work. The band structure is computed by GGA but
a rigid shift of the conduction band (scissor operator) is applied
to fit the band gap to the more accurate value obtained by \textit{GW}.
The color indicates the character of the bands by projections of the
wave function on the different sites and orbitals. Each ternary compound
has one of the red, green or blue color associated with it and the
resulting color is obtained by mixing them in proportion equivalent
to the projections. The red color is always associated with oxygen.
For the only binary B\textsubscript{6}O, we used red for oxygen and
blue for boron. \label{fig:Band-structures-for}}
\end{figure}

We now turn to analyzing the chemical reasons for exceptionally low
effective masses and propose guidelines for future TCO design. Figure
\ref{fig:Band-structures-for} and \ref{fig:Band-structures-for_mixed}
show the band structure of one representative of each of the chemistries
identified (K\textsubscript{2}Sn\textsubscript{2}O\textsubscript{3},
Ca\textsubscript{4}P\textsubscript{2}O, Tl\textsubscript{4}V\textsubscript{2}O\textsubscript{7},
PbTiO\textsubscript{3}, ZrSO, B\textsubscript{6}O and Sb\textsubscript{4}Cl\textsubscript{2}O\textsubscript{5}).
The color scheme indicates the nature of the band obtained by projecting
the wave functions on the different elements. For the ternary compounds,
one of the red, green or blue color is associated with the different
elements and the resulting color is obtained by mixing those primary
colors proportionally to the projections. The red color is always
associated with oxygen. For the only binary B\textsubscript{6}O,
red is used for oxygen and blue for boron. All compounds have dispersive
valence bands, with a high curvature near the valence band maximum
indicative of a low hole effective mass. For all of them, the oxygen
character of the valence band is very mild in agreement with the localized
nature of oxygen 2\textit{p} orbitals. While our candidates cover
different chemistries, the low effective masses valence bands can
be explained by two main mechanisms which are both related to a chemical
way of producing valence bands with low oxygen character. 

The first mechanism leading to compounds with low effective masses
and wide band gaps concerns K\textsubscript{2}Sn\textsubscript{2}O\textsubscript{3},
Rb\textsubscript{2}Sn\textsubscript{2}O\textsubscript{3}, PbTiO\textsubscript{3},
PbZrO\textsubscript{3}, PbHfO\textsubscript{3}, K\textsubscript{2}Pb\textsubscript{2}O\textsubscript{3},
Tl\textsubscript{4}O\textsubscript{3} and Tl\textsubscript{4}V\textsubscript{2}O\textsubscript{7}
and is illustrated in Figure \ref{fig:Band-structures-for}. For these
compounds, the low hole effective mass originates from the hybridization
of the \textit{s} states of a (n-1)\textit{d}\textsuperscript{10}n\textit{s}\textsuperscript{2}
cation (e.g. Sn\textsuperscript{2+}, Pb\textsuperscript{2+}, or
Tl\textsuperscript{1+}) with oxygen 2\textit{p} orbitals. This is
similar to what happens for Cu-based compounds where the closed 3\textit{d}
shell is hybridized with oxygen. However, \textit{s} are significantly
more delocalized than \textit{d} orbitals leading to higher dispersion
of the valence band and lower effective masses.(refs Walsh) So far,
the most studied (n-1)\textit{d\textsuperscript{10} }n\textit{s\textsuperscript{2}}
ion \textit{p}-type oxide has been SnO.\citep{Fortunato2010} However,
SnO suffers from a very small indirect gap (0.7 eV)\citep{Ogo2008}
and its valence band shows anisotropic effective mass due to its layered
structure. Our work shows that going to ternary oxides of Sn\textsuperscript{2+}
can lead to wider band gaps (similarly than going from Cu\textsubscript{2}O
to CuAlO\textsubscript{2} widens the gap of Cu-based compounds) but
also to more isotropic effective masses by modifying the crystal structure.
The Hg\textsubscript{2}SO\textsubscript{4} compound shows also an
orbital overlap of \textit{s} electrons with oxygen \textit{p} but
with a slightly different electronic configuration as Hg\textsuperscript{1+}
is 5\textit{d}\textsuperscript{10}6\textit{s}\textsuperscript{1}.

From our set of (n-1)\textit{d}\textsuperscript{10}n\textit{s}\textsuperscript{2}-containing
compounds, we found that the oxygen hybridization is the most pronounced
for Sn\textsuperscript{2+} followed by Pb\textsuperscript{2+} and
finally Tl\textsuperscript{1+} which has the least dispersive valence
bands and the strongest oxygen character. While Bi\textsuperscript{3+}
is also (n-1)\textit{d\textsuperscript{10} }n\textit{s\textsuperscript{2}
}and some Bi-based compounds were close to meet our criteria (e.g.,
BiVO\textsubscript{4}, Bi\textsubscript{2}Ti\textsubscript{4}O\textsubscript{11},
and BiBrO, see supplementary information), none of them passed the
1.5 cut-off on average effective mass. It is possible that our data
set did not contain the adequate crystal structure to lead to large
bismuth \textit{s}-oxygen \textit{p} overlap in every directions but
that such a structure might be achievable. A Sb\textsuperscript{3+}-based
compound (Sb\textsuperscript{3+} is 4\textit{d\textsuperscript{10}
}5\textit{s\textsuperscript{2}}) is also present in our candidates
(i.e., Sb\textsubscript{4}Cl\textsubscript{2}O\textsubscript{5}).
However, here not only does the \textit{s} orbital of Sb hybridize
with oxygen but the valence band character is also influenced by the
presence of anionic chlorine.

The presence of an additional anion is actually the second mechanism
at play in our low effective mass candidates (i.e., ZrOS, HfOS and
the oxypicnitides) and presented in Figure \ref{fig:Band-structures-for}.
Here, the valence band keeps a \textit{p} character but more delocalized
orbitals such as 3\textit{p} (for S\textsuperscript{2-}, P\textsuperscript{3-})
or 4\textit{p} (As\textsuperscript{3-}) replace the oxygen 2\textit{p}
valence band. Although, pure 3\textit{p} or 4\textit{p}-based compounds
have often too small band gaps, the mixing of oxygen and another \textit{p}-orbital
anion can lead to compounds with low effective masses and large band
gaps. For instance, the band gaps of ZrS\textsubscript{2} is 1.7
eV\citep{Moustafa2009} while ZrO\textsubscript{2} shows a larger
band gap (from 5.2 to 5.7 eV\citep{Gruning2010}) but flat oxygen
valence bands. ZrOS achieves an interesting trade-off offering a larger
band gap than for the sulfide with lower hole effective masses due
to the sulfur character of the valence band. Our study did not lead
to any oxypicnitides offering such a good trade-off but it is a still
unexplored chemistry where future TCO work might be fruitful. Boron
suboxide actually also fits in the anion mixing category as B\textsubscript{6}O
can be seen as a mixture of cationic boron, anionic boron and anionic
oxygen.

From this analysis, two novel design principles can be outlined to
achieve low hole effective masses oxygen containing compounds: the
use of (n-1)\textit{d\textsuperscript{10} }n\textit{s\textsuperscript{2}}
ions especially Sn\textsuperscript{2+} (but also Bi\textsuperscript{3+}
or Sb\textsuperscript{3+}) or the presence of another anion with
\textit{p}-orbitals more delocalized than oxygen (e.g., S\textsuperscript{2-},
P\textsuperscript{3-}, Cl\textsuperscript{-} or Br\textsuperscript{-}).
Both approaches can be combined as in Sb\textsubscript{4}Cl\textsubscript{2}O\textsubscript{5}.
Beyond the compounds we have already identified, there are still many
opportunities for discovering new low hole effective mass TCOs (e.g.,
previously unknown ternaries or quaternaries) with the guidance of
those design principles.

Motivated by the technological need for high mobility \textit{p}-type
TCOs and the difficulty to design such materials, we studied a large
database of computed band structures for thousands of oxides, searching
for the chemistries prone to low effective masses oxides. Our large
scale study confirms that holes tend to have higher effective masses
than electrons in oxides, but identifies several non-toxic, earth-abundant
compounds with exceptionally low hole effective masses and wide band
gaps: B\textsubscript{6}O, Sb\textsubscript{4}Cl\textsubscript{2}O\textsubscript{5},
A\textsubscript{2}Sn\textsubscript{2}O\textsubscript{3}(A = K,
Na) and ZrOS. By analyzing those results, two novel design principles,
leading to compounds with lower effective masses than state of the
art p-type TCOs, emerge: hybridization of a (n-1)\textit{d\textsuperscript{10}
}n\textit{s\textsuperscript{2} }with oxygen and/or mixed anionic
systems. Those principles are extremely valuable to guide future searches
for high mobility \textit{p}-type TCO and, after a decade of exploration
of Cu-based chemistries, they offer new chemical spaces full of promises.

\section*{Methods}

All the high-throughput DFT computations have been performed using
the Vienna ab initio software package (VASP),\citep{Kresse1996} with
PAW pseudopotentials\citep{Blochl1994} and the generalized gradient
approximation (GGA) as implemented by Perdew, Burke and Ernzerhoff
(PBE).\citep{Perdew1996} All ionic relaxations have been performed
using AFLOW\citep{Curtarolo2012} and the high-throughput computations
parameters are described in Jain et al.\citep{Jain}

We considered all compounds containing less than 100 atoms in the
unit cell present in the Materials Project\citep{MP} Database originally
from the Inorganic Crystal Structure Database (ICSD) and containing
oxygen and no elements such as rare-earths (from Z=58, Ce to Z=71,
Lu), inert gases, and any element with an atomic number larger than
84 (Po).

To perform the high-throughput band structure computations, we used
the Materials Project structures (already relaxed with GGA) and performed
a static run to obtain the charge density followed by a non-self-consistent
band structure run along the band structure symmetry lines provided
by Curtarolo \textit{et al}.\citep{Setyawan2010,Curtarolo2012} The
full average effective mass tensor computation was performed on a
regular gamma centered 8,000 \textbf{k}-points grid interpolated using
the Boltztrap code.\citep{Madsen2006}

In the one-shot \textit{GW} approach, corrections are obtained perturbatively
from a starting DFT electronic structure. \textit{GW} and preparatory
DFT calculations on the 20 target compounds are performed with the
ABINIT code\citep{Gonze2009} at optimized geometries, obtained from
the Materials Project database. The exchange correlation energy for
the preparatory DFT computation is described using the local density
approximation (LDA) functional.\citep{Perdew1992} The Brillouin Zone
in sampled with Monkhorst-Pack grids and the k-point sampling density
is similar for all considered systems (> 450/\textit{n} \textbf{k}-points
where \textit{n} is the number of atoms in the unit cell). For each
oxide, the planewave cutoff is determined separately and set using
a total energy difference convergence criterion, leading to electronic
energies converged within 10\textsuperscript{-3} eV on average. We
use norm-conserving pseudopotentials to model the electron-ion interaction
(see supplementary information). If any, we include semi-core \textit{d}
states as valence in the pseudopotential. The \textit{GW} calculations
are carried out using the the well established Godby-Needs plasmon
pole approximation.\citep{Godby1989} We use a cutoff of 20 Rydberg
for the expansion of the dielectric matrix and a total number of $\sim$1300
bands for all oxides .

All analysis of the data (e.g., line effective mass computations or
band structure plotting) was performed using the pymatgen python package.\citep{Ong2012}

\section*{References }

\bibliographystyle{naturemag}

\section*{Acknowledgments}

Geoffroy Hautier acknowledges the F.N.R.S.-FRS for financial support
through a ``chargé de recherche'' grant. Gerbrand Ceder acknowledges
the ONR (Office or Naval Research) for financial support through the
award N00014-11-1-0212 of ``New Technologies Through Computational
Materials Design''. The authors acknowledge technical support and
scientific advice from J.-M. Beuken, M. Giantomassi, M. Stankovski,
D. Waroquiers, Y. Pouillon, M. Kocher, A. Jain, S. P. Ong, and K.
Persson. This work was also supported by the Agentschap voor Innovatie
door Wetenschap en Technologie (IWT project N\textdegree{}080023,
ISIMADE) , the FRS-FNRS through FRFC project 2.4.589.09.F, and the
Région Wallonne through WALL-ETSF (project Number 816849). Computational
resources have been provided by the supercomputing facilities of the
Université catholique de Louvain (CISM/UCL) and the Consortium des
Équipements de Calcul Intensif en Fédération Wallonie Bruxelles (CÉCI)
funded by the Fond de la Recherche Scientifique de Belgique (FRS-FNRS).
The authors would like to thank MIT-UCL MISTI global seed fund for
financial support.
\end{document}